\documentstyle[12pt,amsfonts,amssymb,amscd]{article}
\title{Black Hole Entropy, Topological Entropy and the Baum-Connes Conjecture
in K-Theory}
\author{Ioannis P. \ ZOIS\thanks{izois\,@\,maths.ox.ac.uk; Research supported by the EU, contract no HPMF-CT-1999-00088}\\
\\
Mathematical Institute, Oxford University
\\
24-29 St. Giles', Oxford OX1 3LB.}
\date{}

\begin{document}
\maketitle
\begin{abstract}
We shall try to exhibit a relation between \emph{black hole} entropy and
\emph{topological entropy} using the famous {\bf Baum-Connes conjecture} for
\emph{foliated manifolds} which are particular examples of
\emph{noncommutative spaces}. Our argument is \textsl{qualitative} and it is
based on the microscopic origin
of the Beckenstein-Hawking area-entropy formula for black holes, provided by
superstring theory, in the more
general noncommutative geometric context of M-Theory following the Connes-
Douglas-Schwarz article.\\

PACS classification: 11.10.-z; 11.15.-q; 11.30.-Ly\\

Keywords: Godbillon-Vey class, String Theory, Foliations, Dynamical Systems,
Black Holes, Topological Entropy.\\
\end{abstract}

\section{Introduction and motivation}

We know from a series of articles back in 1996
due to Strominger, Vafa, Maldacena and Horowitz \cite{S} that superstring
theory can
in some cases (\textsl{multicharged extremal black holes} and for large values
of charges) give an explanation for the
\textsl{microscopic origin} of the quantum states associated to a black hole,
which give rise to its quantum mechanical entropy described by the
Beckenstein-Hawking area-entropy formula.

The argument relies heavily on \emph{S-duality} which
gives a way to \textsl{identify}
\textsl{perturbative string states and $D$-branes}, these are all $BPS$
states, in \emph{weak coupling} region with \textsl{extremal black holes
with $NS$ and $R$ charges} respectively in \emph{strong coupling} region.

 A crucial detail to bear in mind is that
since superstring theory lives in 10 dimensions and the Beckenstein-Hawking
formula refers (originally) to 4 dimensions, the extra dimensions have to be
\emph{compactified}; hence compactification is important in establishing
this relation.\\

In 1998 the now ``classical'' article due to A. Connes, R. Douglas and A.
Schwarz \cite{CDS} tought us that \emph{M-Theory}, which is a
\textsl{generalisation}  of superstring theory, admits  \emph{additional}
compactifications on \emph{noncommutative spaces},
in particular \textsl{noncommutative tori}.\\

Then the natural question is:\\

What would happen if in the scenario considered by Strominger, Vafa
\textsl{et al.}, we now assume that the \emph{compactified dimensions} form a
\emph{noncommutative space}?\\

We shall try to give a \emph{qualitative answer} to the above question mainly
based on \emph{(noncommutative) topology}.\\

Before doing that, we shall make some brief remarks on both M-Theory and
Noncommutative Geometry.\\

We start with M-Theory: until mid 90's we had 5 consistent superstring
theories: Types I, IIA, IIB, heterotic $SO(32)$ and heterotic
$E_{8}\times E_{8}$.
After the discoveries of various string dualities, it is now believed that
these 5 theories are an artifact of perturbation expansion: there is only
\emph{one fundamental 11-dim theory} called \emph{M-Theory} which contains
$p$-dimensional extended objects called \emph{$p$-branes}. For example,
point particles are 0-branes, strings are 1-branes etc.
Rather few things are known about this
underlying theory and the basic strategy is to try to understand this M-Theory
from its \emph{limiting theories} which are the 5 superstring theories in 10
dimensions and 11 dimensional supergravity.

Next we shall try to give an idea of what
\emph{noncommutative geometry} is. The motivation for the development of this
\textsl{new branch of mathematics} is actually 2-fold:

{\bf 1.} \emph{Descartes} introduced \emph{coordinates} in the 17th century
and
revolutionised geometry. Subsequently that gave rise to the notion of
\emph{manifold}. One important generalisation introduced by Alain Connes
(see \cite{Connes}) was the notion of a \emph{noncommutative manifold}.
Roughly, one can think of a ``generalised manifold'', or
``noncommutative manifold'', as a space having a corresponding function
space which locally
``looks like'' an \emph{operator algebra}, in fact a $C^{*}$-algebra which
in general is \textsl{noncommutative}
instead of just functions on some Euclidean space ${\bf R^{n}}$ which is the definition
for an ordinary manifold as we know it from geometry. This is strongly
reminiscent of quantum mechanics and sometimes these are called
\textsl{``quantum spaces''}. The origin is essentially
\emph{Gelfand's theorem} which states that the \textsl{category} of (unital)
\emph{commutative} $C^{*}$-algebras with $*$-preserving homomorphisms is
equivalent
to the \textsl{category} of (compact) \emph{locally compact Hausdorff spaces}
with homeomorphisms.\\

{\bf 2.} We would like to generalise the \emph{Index Problem} solved by
Atiyah and Grothendieck in late 60's. The origin came from \emph{Quillen's
Higher Algebraic K-Theory}, a \textsl{simplification} of which is the
\emph{K-Theory of} (not neccessarily commutative) \emph{$C^{*}$-algebras}
which we shall use later.
Then \emph{Serre-Swan theorem} identifies it with Atiyah's original K-Theory
in the \textsl{commutative case} using Gelfand's theorem.

We think that the idea behind the first motivation is quite clear and in fact
\emph{this} idea is behind the vast majority of articles in physics literature up to
now which make some use of noncommutative geometry. We shall
not give the precise definitions here. The interested reader may study
\cite{Connes} which also contains an exhaustive list of references on the
subject.

In this article we would like however to elaborate
more on the ideas behind the \textsl{second motivation}, namely
\emph{Index Theory;} in fact one of the aims of
this present article is to try to make some use of the ideas behind it in
physics and we shall start by explaining
what \emph{Index Theory} is (we have been influenced in our presentation by
\cite{BD} which is an excellent article).\\

Index theory is an attempt to unify \emph{topology} and \emph{analysis}.\\

The formal way to do that is to manufacture two mathematical objects, one
containing
the topological data and the other containing the analytical data and then we
compare them;
more concretely, given a ``commutative'' space $M$ (namely a manifold or
an algebraic variety), one constructs \emph{two} \textsl{K-Theories}: one is
called \emph{topological} and contains all stable isomorphism classes
of (say) complex vector bundles over the space $M$.
The other is called \emph{analytical} but we shall adopt the
more recent term \emph{K-Homology} and contains all
homotopy classes of principal symbols of elliptic pseudodifferential
operators acting on $M$ (more precisely on sections of vector bundles
over $M$). What we describe is Atiyah's $Ell$ group from which K-Homology
evolved subsequently.

Grothendieck proved that for any commutative space the analytical and the
topological K-Theories are \emph{isomorphic} and then one can say that
essentially the \emph{Atiyah-Singer Index Theorem} gives the
\emph{explicit isomorphism}.

One also has two natural maps from these two K-Theories to the integers:
for the \emph{topological K-Theory} it is given by the
\textsl{Chern character} and for \emph{K-Homology} it is
given by the \textsl{(Fredholm) Index} of the operator. Then the
Atiyah-Singer Index theorem says that the Index differs from the Chern
character essentially by the \emph{Todd class}.\\

{\bf Remark 1:} The relation between topology and analysis is quite deep; the
Atiyah-Singer Index Theorem gives a relation between \emph{primary}
invariants (Chern classes and Index).
There are also relations between \emph{secondary invariants}, which are more
\textsl{delicate objects} like
\emph{Chern-Simons} forms for bundles and Atiyah's intriguing
\emph{$\eta $ invariant} for operators (related to Riemann's famous
``zeta'' function). The \emph{Jones-Witten topological quantum field theory}
on 3-manifolds is such an example, if one thinks of it as the \emph{non-Abelian
version} of A. Schwarz's original work where he observed that there is a close
relation between the partition function of Abelian Chern-Simons 3-form (degenerate quadratic functionals) and
the \emph{Ray-Singer analytic torsion} (the $\eta $ invariant of the
Laplacian) which is a topological invariant of the
3-manifold considered (see \cite{sch}).\\

{\bf Remark 2:} Each of the above two K-theories essentially consists of 2
Abelian groups due to Bott periodicity, namely we have topological
$K^{0}(M)$ and $K^{1}(M)$ and
analytical $K_{0}(M)$ and $K_{1}(M)$, where in the later we have put the
indices downstairs to indicate that this is a homology theory (K-Homology).\\

The \emph{Baum-Connes conjecture} then is an analogous generalised statement
for \emph{analytical}
and \emph{topological} K-Theories \textsl{appropriately defined} for
\textsl{noncommutative spaces};
in fact in its most general formulation it refers to categories with
inverses (groupoids).

We shall only mention here that the basic tool to construct these K-Theories
for categories is essentially
the \emph{Quillen-Segal} construction (see for example \cite{Z1} and references
therein).

\section{Microscopic Origin of Black Hole Entropy}

We shall treat the simplest
example appearing in \cite{S} (we use the shorthand notation ``BH''
for black holes; see moreover \cite{das} which is a nice review article on
the subject):

 Consider a 5-dim BH with 3 charges $Q_{1}, Q_{5}, n$. Since
superstrings require 10 dimensions, we assume the remaining 5 dims are
\emph{compactified} on a fixed torus of volume $(2\pi )^{4}V$ which is
constant
and the 5th remaining direction is another circle of circumference $2\pi R$,
where this radius is much bigger than those of the other 4 circles in the
4-torus.
One can compute using BH quantum mechanics that
$$S_{BH}=\frac{A}{4G}=2\pi \sqrt{Q_{1}Q_{5}n}$$
The same result can be obtained from string theory considerations:
apart from the metric, one has an NS field $H$ (3-form) with both electric
and magnetic charges denoted $Q_{1}$, $Q_{5}$ and $n$ is the quantization
of the momentum $P=n/R$ along
the large circle. If we assume type IIB superstring theory and start from flat
10-dim spacetime we compactify on the 5-torus as described above. The objects
which carry the
charges $Q_{1}$ and $Q_{5}$ turn out to be respectively a $D$-string wrapped
$Q_{1}$ times around the big circle of radius $R$ and a $D5$-brane wrapped
$Q_{5}$ times around the 5 torus. We would like to underline here that the
calculation appearing in \cite{S} is an Index Theoretic one because what
the authors use in order to count BPS states is the \textsl{supersymmetric
Index}.\\

Then our question which we mentioned in the first section was to see how
this formula should be modified if we assume that the compactified 5-torus
is a \emph{noncommutative} one. In addition we shall also assume that the
noncommutative 5-torus
is an ordinary 5-torus which carries a \emph{foliation structure}. The reason
for this is that the \emph{spaces of leaves} of foliations can be really
``very
nasty spaces'' from the topological point of view and in most cases they are
not (ordinary) manifolds. So foliated manifolds are particular examples of
noncommutative manifolds. More details and examples can be found in
\cite{Connes}.

{\bf Suggestion:}

The difference will be in the topological charge $Q_{5}$. We should use an
\emph{invariant} for \textsl{foliated manifolds}. Our suggestion is the
\textsl{new} invariant introduced in \cite{Z2} coming from the pairing
between \emph{K-Homology} and \emph{cyclic cohomology}. The formula is:

$$<[e],[\phi ]>=(m!)^{-1}(\phi \# Tr)(e,...,e)$$

where $e\in K_{0}(C(F))$, $\phi \in HC^{2m}(C(F))$ and $\# $ is the
\emph{cup product} in cyclic cohomology introduced by Connes. In the above
formula we denote by $F$ the codim-$m$ foliation of the 5-torus, $C(F)$
is the $C^{*}$-algebra associated to the foliation (which comes after
imposing a
suitable \textsl{$C^{*}$-algebra ``completion''} to
the \emph{holonomy groupoid} of the foliation) and finally $[e]$
and $[\phi ]$
are \emph{``canonical'' classes} associated to the foliation. The first one
is a naturally chosen \emph{closed transversal} and the second is the
\emph{fundamental cyclic cocycle} of the \emph{normal bundle} of the foliation.
Moreover $K_{0}(C(F))$ and $HC^{2m}(C(F))$ denote the 0th K-Homology group
and the $2m$-th cyclic cohomology group of the corresponding $C^{*}$-algebra
of the foliation respectively. (More details and precise definitions can be
found in \cite{Z2}).\\

The definition of the above invariant uses K-Homology, namely it is
{\bf operator algebraic}. That means that it lies in the \emph{analytical
world}. (The above framework uses the language of $C^{*}$-algebras which by
definition is a combination of algebra and functional analysis). \textsl{We
would like to see what it corresponds to in the}
\emph{topological world}. This would have been very straightforward if we
had known that the Baum-Connes conjecture was true.\\

Last year a \emph{deep} theorem was proved by G. Duminy and
that refers to foliated manifolds as well but it uses
{\bf topological tools} hence it lies in the \emph{topological world}.
It is very
interesting to try to see \textsl{how it is related to our invariant}. We have
gained some better, at least qualitatively, understanding of this relation
\cite{connes1}:

 This invariant is an \emph{integer} since it comes as
 the \emph{Fredholm Index} of some leafwise elliptic operator (see \cite{Connes}, the index theorem due to Connes-Moscovici).\\

Note that an important property of this invariant is that in the
\textsl{commutative case} namely for a fibre bundle, it does not vanish as
the GV-class does
(recall that the GV-class is a particular class in the Gelfand-Fuchs
cohomology) but it reduces to
the usual characteristic classes (linear combination of the Chern class of
the bundle which is the foliation itself,plus the Pontrjagin class of
the tangent bundle of the base manifold which in this case is the normal
bundle of the foliation, see \cite{Z2}).\\

Based on the above commutative example, a \textsl{qualitative picture} is
that in the general case of an arbitrary foliation, this
invariant is the sum of two parts:
the first is some Chern (or Pontrjagin) class of the normal bundle of our
foliation and
the second is some characteristic class of our foliation itself, namely a
class of the corresponding Gelfand-Fuchs cohomology. Moreover we know from
the Duminy
theorem that (for codim-1 cases) the GV-class is related to the topological
entropy and thus the second, noncommutative part of our invariant, should
``contain'' the difference in the entropy. \\

Essentially
what we are trying to do is to understand some of the mysteries of the
Baum-Connes conjecture in the
particular case of foliated manifolds. We have not succeded in doing this but
we think it is worth reviewing the topological side of the story along
with Duminy's theorem. Needless to say that the Baum-Connes conjecture is one
of the major mathematical problems still open today which attracts a lot of
interest from pure mathematicians.

\section{Duminy's Theorem}

Up to a large extend, what we know for the topology of foliated manifolds,
is essentially due to the
pioneering work of W. Thurston in late '70's and it refers primarily to
codim-1 foliations on closed 3-manifolds.

There is only one known invariant for foliated manifolds,
which is roughly the analogue of the Chern classes for bundles: this is the
celebrated \emph{Godbillon-Vey class} which belongs to the
\textsl{Gelfand-Fuchs cohomology}.

Let us review some basic facts for \emph{foliated manifolds;} roughly they
generalise fibre bundles
(the total space of every fibre bundle is a foliation, the fibres are the
leaves):

By definition a codim-$q$ \emph{foliation} $F$ on an $m$-manifold $M$ is
given by a
codim-$q$ \emph{integrable subbundle} $F$ of the tangent bundle $TM$ of $M$.
``Integrable'' means that the Lie bracket of vector fields of $F$ closes.
This is the \emph{global} definition of a foliation.

 There is an equivalent \emph{local} definition: a codim-1 foliation $F$
on a smooth
$m$-manifold $M$ can be defined by a non-singular 1-form $\omega $ vanishing
exactly at vectors tangent to the leaves. Integrability of the corresponding
$(m-1)$-plane bundle $F$ of $TM$ implies that $\omega \wedge d\omega =0$ or
equivalently $d\omega =\omega \wedge \theta $ where $\theta $ is another
1-form. The 3-form $\theta \wedge d\theta $ is closed hence determines a de
Rham cohomology class called the \emph{Godbillon-Vey} class of $F$
(abreviated to ``GV'' in the sequel).

Although $\omega $ is only determined by $F$ up to multiplication by nowhere vanishing functions and $\theta $ is determined by $\omega $ only up to addition of a $d$-exact form, actually the Godbillon-Vey class depends only on the foliation $F$. The Godbillon-Vey class can also be defined for foliations of codim grater than 1 and $\theta $ can be thought of as a \emph{basic} (or sometimes called \emph{Bott}) connection on the normal bundle of the foliation which by definition is $TM/F$ (see \cite{Z2} and references therein). For a codim-$q$ foliation the GV class is a $(2q+1)$-form.

Note that following the global definition of a foliation given above, the
subbundle $F$ of the tangent bundle $TM$ of $M$ is itself an honest bundle
over $M$ and thus it has its own characteristic classes from Chern-Weil theory.
This theory however is \emph{unable} to detect the \textsl{integrability
property} of $F$ and for this reason we had to develop the Gelfand-Fuchs
cohomology, a member of which is the GV-class.\\

The \emph{key thing} to understand about foliations is that a codim-$q$
foliation $F$ on an $m$-manifold $M$ gives a \emph{decomposition} of
$M$ into a \textsl{disjoint union} of submanifolds called
\emph{leaves} all of which have the \emph{same}
dimension $(m-q)$. The definition of a foliation seems rather
``innocent'', at least the global one, maybe because it is very brief. Yet
this is very far from being true. One has {\bf two} \textsl{fundamental
differences} between a foliation and the \textsl{total space} of a fibre
bundle:\\

{\bf 1.} The leaves of a foliation in general have \emph{different fundamental
groups} whereas for a bundle the
fibres are the \emph{``same''} (homeomorphic) as some fixed space called
\emph{typical fibre}. Thus genericaly one has no control on the homotopy types
of the leaves; under some very special assumptions however (e.g. restrictions
on the homology groups of the manifold which carries the foliation) one may
get ``some''
control on the homotopy types of the leaves and in these cases we obtain some
deep and powerful theorems, the so called \emph{stability theorems}.

The above fact, along with the \emph{holonomy groupoid} of the foliation
(roughly
the analogue of the \textsl{group of gauge transformations} for principal
bundles) give
rise to a corresponding \emph{noncommutative algebra} which one can naturally
associate to any foliation using a construction due to A. Connes; for
fibrations the corresponding algebra is essentially \emph{commutative.}
(``essentially'' means it is \emph{Morita Equivalent} to a commutative one;
 for the proof see \cite{Z2}). Moreover some leaves may be compact
and some others may not.\\

{\bf 2.} The leaves are in general \emph{immersed} submanifolds and
{\bf not} \emph{embedded} as the fibres
of a fibration. In both cases normally there is no
intersection among different leaves and fibres (we assume for simplicity no
singularities) so in both cases one can say that we have a notion of
\emph{parallelism}. For foliations
it is \textsl{far more general}; that can give rise to
\emph{topological entropy.} This notion was introduced by topologists
 (Ghys, Langevin and Walczak) in 1988 (see \cite{ghys} or \cite{CC}).\\

We need one further definition before we state Duminy's theorem:
 A leaf $L$ of a codim-1 foliation $F$ is called \emph{resilient} if
there exists a transverse arc $J=[x,y)$ where $x\in L$ and a loop $s$ on $L$
based on $x$ such that $h_{s}:[x,y)\rightarrow [x,y)$ is a contraction to $x$ and the intersection of $L$ and $(x,y)$ is non-empty.  (Note that in the definition above the arc $J$ is \textsl{transverse} to the foliation). Intuitively a resilient leaf is one that ``captures itself by a holonomy contraction''. The terminology comes from the French word \emph{``ressort''} which means
\textsl{``spring-like''}. We are now ready to state

 {\bf Duminy's Theorem:}

\textsl{``For a codim-1 foliation $F$ on a closed smooth $m$-manifold $M$ one
has that $GV(F)=0$ unless $F$ has some (at least one)} \emph{resilient}
leaves''.\\

The \textsl{proof} is very long and complicated and it uses a theory called
\emph{architecture of foliations} (see \cite{CC}).\\

The important lesson from G. Duminy is that for topology, \emph{only
resilient leaves matter}, since only them contribute to the GV-class.

As a very interesting Corollary of the above theorem we get the relation
between the GV-class and
\emph{topological entropy}. To define this notion one has first to
define the notion of \emph{entropy of maps} and then generalise it for
foliations using as intermediate steps the entropy of transformation groups
and pseudogroups.

 In general, \textsl{entropy measures the rate of creation of information}. Roughly, if the states of a system are described by iteration of a map, states that may be \textsl{indistinguishable} at some initial time may diverge into clearly \textsl{different} states as time passes. Entropy measures the \emph{rate} of creation of states. In the mathematical language it measures the
\textsl{rate of divergence of orbits of a map}.

We shall give a qualitative description:
Let $f$ be a map from a compact manifold onto itself. To measure the number of orbits one takes an empirical approach, not distinguishing $\varepsilon $-close points for a given $\varepsilon >0$. If $x$ and $y$ are two indistinguishable points, then their orbits $\{f^{k}(x)\}_{k=1}^{\infty }$ and $\{f^{k}(y)\}_{k=1}^{\infty }$ will be distinguishable provided that for some $k$, the points $f^{k}(x)$ and $f^{k}(y)$ are at distance grater than $\varepsilon $. Then one counts the number of distinguishable orbit segments of length $n$ for fixed magnitude $\varepsilon $ and looks at the growth rate of this function of $n$. Finally one improves the resolution arbitrarily well by letting $\varepsilon \rightarrow 0$. The value obtained is called \emph{the entropy
of $f$} and it \emph{measures} the \textsl{asymptotic growth rate of the
number of orbits of finite length as the length goes to infinity}.

The above can be rigorously formulated and one can define the \emph{entropy of a foliation} to be a \textsl{non-negative real number} (see \cite{ghys}).

One then can prove:

{\bf Proposition:}\\
If the compact foliated space $(M,F)$ has a \emph{resilient} leaf, then $F$ has \emph{positive} entropy.\\

The proof can be found in \cite{CC}.

Combining this with Duminy's theorem (for \emph{codim-1 case}) we get the following\\

{\bf Corollary:}\\
If $(M,F)$ is a compact ($C^2$-)foliated manifold of \textsl{codim-1}, then
\emph{zero entropy implies GV(F)=0}.\\

\section{Physical discussion:}

Topologically, the difference between the commutative charge and the
noncommutative one
is the \emph{topological entropy} of the
foliated torus.
Commutative spaces can be considered to have \emph{zero} topological entropy
whereas foliations \textsl{may} have non-zero topological entropy.

{\bf Note:}
\emph{Not every noncommutative space has non-zero topological entropy}.
Duminy's theorem tells us that this is ``captured'' by the GV-class.

Physically, one can try to think of some ``critical point'' where the foliation
becomes ``wild enough'' in order to develop resilient leaves, thus have
non-zero GV-class and thus non-zero topological entropy. Geometrically the parameter which indicates the transition from the commutative to the noncommutative realm is exactly the GV class since
it is the parameter which signifies the appearence of non-zero
topological entropy. It would be very interesting to try to see if the GV class has any direct physical meaning: one suggestion would be that it might be related to the
\emph{curvature of the $B$-field} for the \textsl{codim-1 case} in some appropriate context (see \cite{Z3} for more details).

Moreover it is very desirable from the physical point of view to try to find a
\textsl{quantitative} description
of this scenario via a \emph{direct} computation using \emph{(almost)BPS}
states.
Some recent work (mainly last year) due to
Konechny and Schwarz \cite{KS} might be useful in this direction.
Let us fix our notation: $T^{d}$ denotes the \emph{commutative} $d$-torus and $T^{d}_{\theta }$ denotes \emph{noncommutative} $d$-torus.
Of particular interest
is the case of \emph{noncommutative} ${\bf Z_{2}}$ and ${\bf Z_{4}}$
 \emph{toroidal orbifolds}
considered by Konechny-Schwarz in their most recent articles.\\

The role of supersymmetry is very important: our understanding is that
supersymmetry prevents the foliation
from becoming ``very messy'' in order to have non-zero GV class.
Supersymmetry and topological entropy are mutually \emph{``competing''}
notions. We would like to find \emph{how much supersymmetry is needed to be
preserved so that the topological entropy remains zero}.

For example in
all the cases considered in the Connes-Douglas-Schwarz article \cite{CDS}, the
foliations of the tori were \emph{linear} (\textsl{Kronecker foliations} as
they are known in geometry), so topologically they were
spaces with zero topological entropy. That was dictated by their
\textsl{maximal supersymmetry} assumption ({\bf constant} 3-form field
$C$ in their D=11 supergravity interpretation). In most cases studied up to
now in physics literature this is also the case. In the recent
articles by Konechny-Schwarz however quoted above, this is probably no longer the
case. For the case of $K3$ for example which can be described as an orbifold $T^{4}/\Gamma $, where $\Gamma $ any discrete group, considering orbifolds corresponds to breaking \emph{half} of the supersymmetry.
Konechny and Schwarz studied the \emph{moduli space} of \textsl{constant curvature
connections}
on \textsl{finitely generated projective modules} (this should be thought of
as the noncommutative
analogue of fibre bundles) over algebras of the form (we follow their
notation) $B^{d}_{\theta }:=T^{d}_{\theta }\rtimes {\bf Z}_{2}$, where
$T^{d}_{\theta }$ is our friend the noncommutative $d$-dimensional torus. Let us denote by  $B^{d}:=T^{d}\rtimes {\bf Z}_{2}$ the commutative ${\bf Z}_{2}$
toroidal orbifold (when we write $T^{d}$ we mean functions on $T^{d}$ to be absolutely precise but by Gelfand's theorem these are identified).
These connections correspond to \emph{$\frac{1}{2}$ BPS states}. Then the
volume of the
moduli space is related to the number of quantum states by standard physical
arguments. The first question is: does the foliation corresponding to the algebra $B^{d}_{\theta }$ have non-zero GV-class? If yes, our topological discusson is of much interest, if not one should break more supersymmetry in order to make noncommutative topological phenomena appearing. More work is
certainly needed in order to understand these \emph{fractional} BPS states from the physics point of view.\\

Our ideas seem to be supported by two observations, the first one is made
in \cite{KS}:

1. When the authors in \cite{KS} tried to count $1/4$ BPS states on the noncommutative 3-torus $T^{3}_{\theta }$
they observed that the result agreed with the result obtained in \cite{HV}
for the \emph{commutative} 3-torus $T^{3}$. This means that the noncommutative torus alone is not enough for noncommutative topology.

2. The 0th K-Theory group of the ${\bf Z_{2}}$ noncommutative toroidal
orbifold $B^{d}_{\theta }$ is the \emph{same} as the commutative ${\bf Z_{2}}$ toroidal orbifold $B^{d}$ which in turn is the same as the ${\bf Z_{2}}$-equivariant K-Theory of $T^{d}$. More concretely
$$K_{0}(B^{d}_{\theta})\cong K_{0}(B^{d}):=K_{0}(T^{d}\rtimes {\bf Z}_{2})\cong
K^{0}_{Z_{2}}(T^{d})={\bf Z}^{3\cdotp 2^{d-1}}.$$
The above result follows from the work of Julg and Walters  \cite{J} and
\cite{Wal}.\\

So to \emph{conclude}, in this article
we argued that the assumption that the compactified dimensions form a
noncommutative torus \emph{will have consequences} for the black hole
area-entropy formula, provided that the foliated torus is
\emph{``messy enough''} to have resilient leaves. Our argument was purely
topological.

Let us close with the following remark: in all these articles \cite{KS} there are no \emph{cyclic (co)homology} groups
appearing, the reason \emph{possibly} being that \emph{topologically} these
spaces are in
fact commutative (tori which can be continously deformed to the commutative
case where the noncommutativity parameter $\theta $ is zero), despite the
fact that they are called noncommutative.

Our discussion was about foliated manifolds (tori in particular) which have
indeed extra noncommutative topological charges, namely either the GV class
or our new operator algebraic invariant which uses cyclic (co)homology.

Moreover, since it is very important for string theory
to understand some nonsupersymmetric background, it is perhaps the case that
as far as noncommutative geometry is concerned, in order to have some
nontrivial topological phenomena appearing (e.g. nonzero topological entropy),
one must brake supersymmetry completely. This suggests that an
understanding of nonsupersymmetric string vacua may give some better
understanding of the Baum-Connes conjecture at least for the particular
case of foliated manifolds and vice-versa, namely if one wants to understand nonsupersymmetric string vacua one must use noncommutative topology. That was the second point we tried
to argue here and stimulate research both from mathematics and from physics.


\begin{thebibliography}{20}




\bibitem{sch}A.S. Schwarz: ``The partition function of degenerate quadratic
functional and the Ray-Singer invariants'', Lett. Math. Phys. 2 (1978).\\

\bibitem{Z2}Zois, I.P.: ``A new invariant for $\sigma $ models'', Commun. Math. Phys. Vol 209 No 3 (2000) pp757-783.\\

\bibitem{Z1}Zois, I.P.: ``The Godbillon-Vey class, Invariants of Manifolds
and Linearised M-Theory'', hep-th/0006169, Oxford preprint, (submitted to
Commun. Math. Phys., Prof. A. Connes).\\

\bibitem{Z3}Zois, I.P.: ``Black Hole Entropy, Topological Entropy and
Noncommutative Geometry'', hep-th/0104004, Oxford preprint.\\

\bibitem{CC}A. Candel and L. Conlon: ``Foliations I'', Graduate Studies in Mathematics Vol 23, American Mathematical Society, Providence Rhode Island 2000.\\

\bibitem{Connes}A. Connes: ``Non-commutative Geometry'', Academic Press 1994.\\

A. Connes: ``Noncommutative Geometry Year 2000'', qa/0011193.\\

\bibitem{connes1}Private communication with Alain Connes during the DMV-Seminar
on Noncommutative Geometry, Mathematisches Forschunginstitute Oberwolfach,
Germany, 14-20 October 2001.\\

\bibitem{CDS}A. Connes, M.R. Douglas and A. Schwarz: ``Noncommutative Geometry and M-Theory: compactification on tori'', JHEP 02, 003 (1998).\\

\bibitem{BD}P. Baum and R. Douglas: ``K-Homology and Index Theory'', Proc.
Sympos. Pure Math. 38, Providence, Rhode Island AMS 1982.\\

\bibitem{ghys}E. Ghys, R. Langevin and P. Walczak: ``Entropie geometrique des feuilletages'', Acta Math. 160 (1988), 105-142.\\

\bibitem{S}A. Strominger, C. Vafa: ``Microscopic Origin of the Beckenstein-
Hawking Entropy'', Phys. Lett. B379 (1996) 99-104.\\

G. Horowitz, A. Strominger: ``Counting States of Near-Extremal Black Holes'',
Phys.Rev.Lett. 77 (1996) 2368-2371.\\

G. Horowitz, J. Maldacena, A. Strominger: ``Nonextremal Black Hole
Microstates and U-duality'', Phys.Lett. B383 (1996) 151-159  .\\

G.T. Horowitz: ``The Origin of Black Hole Entropy in String
Theory'', Proceedings of the Pacific Conference on Gravitation and Cosmology,
Seoul, Korea, February 1-6, 1996.\\

\bibitem{KS}A. Konechny and A. Schwarz: ``Moduli spaces of maximally
supersymmetric solutions on noncommutative tori and noncommutative orbifolds'',
JHEP 0009 (2000) 005.\\

A. Konechny and A. Schwarz: ``Compactification of M-Theory on
noncommutative toroidal orbifolds'', Nucl.Phys. B591 (2000) 667-684.\\

A. Konechny and A. Schwarz: ``1/4-BPS states on noncommutative tori'',
JHEP 9909 (1999) 030.\\

A. Konechny and A. Schwarz:  ``Supersymmetry algebra and BPS states of super
Yang-Mills theories on noncommutative tori'', Phys.Lett. B453 (1999) 23-29.\\

\bibitem{HV}F. Hacquebord and H. Verlinde: ``Duality Symmetry of N=4
Yang-Mills on $T^{3}$'', Nucl. Phys. B508 (1997) 609-622.\\

\bibitem{J}P. Julg: ``K-theorie equivariante et produits croises'', C.R.
Acad. Sci. Paris, ser. I. Math. 292 (1981), no 13, 629-639.\\

\bibitem{Wal}S.G. Walters: ``Projective Modules over the noncommutative sphere'', J. London Math. Soc. (2) 51 (1995) 589-602.\\

S.G. Walters: ``Chern Characters of Furier modules'', Can. J. Math. (1999), 39 pages (to appear).\\

S.G. Walters: ``K-Theory of noncommutative spheres arising from the Fourier automorphism'', preprint (1999).\\

\bibitem{das}S.R. Das and S.D. Mathur: ``The Quantim Physics of Black Holes:
Results from String Theory'', Ann. Rev. of Nucl. and Particle Science Vol 50
 (2000), gr-qc/0105063.\\

\bibitem{Town}P.K. Townsend: ``Black Holes'', Lecture Notes, DAMTP Cambridge
University, 1998.\\


\end{thebibliography}
\end{document}